\documentclass{elsart}

\usepackage{graphicx}
\usepackage[]{float}

\newcommand{\halfwidth}{0.495\textwidth}
\newcommand{\medwidth}{0.8\textwidth}

\newcommand{\be}{\begin{equation}}
\newcommand{\ee}{\end{equation}}
\newcommand{\bea}{\begin{eqnarray}}
\newcommand{\eea}{\end{eqnarray}}
\newcommand{\bml}{\begin{mathletters}\baselineskip 10pt}
\newcommand{\eml}{\baselineskip 12pt \end{mathletters}}

\newcommand{\m}{{\scriptscriptstyle -}}
\newcommand{\p}{{\scriptscriptstyle +}}

\newcommand{\bra}{\langle}
\newcommand{\ket}{\rangle}
\newcommand{\cond}{\bra 0 | \bar \psi \psi | 0 \ket}

\newcommand{\simleq}{\scriptstyle{\stackrel{<}{\sim}}}

\newcommand{\vc}[1]{\mbox{\bf #1}}

\newcommand{\vcg}[1]{\mbox{\boldmath$#1$}}

\begin{document}

%remove for camera-ready copy
%\begin{flushright}TPI 00/xx \\ August 2000 \end{flushright}

\begin{frontmatter}

%\hyphenation{ ... } 

\title{The Light-Cone Wave Function of the Pion}

\author{T. Heinzl}

%remove footnote for camera-ready copy

\address{Theoretisch-Physikalisches Institut \\
Friedrich-Schiller-Universit\"at Jena \\
Max-Wien-Platz 1 \\
D-07743 Jena}

\date{14 August 2000}

\begin{abstract}
The light--cone wave function of the pion is calculated within the
Nambu--Jona-Lasinio model. The result is used to derive the pion
electromagnetic form factor, charge radius, structure function,
$\pi$--$\gamma$--transition form factor and distribution
amplitude. 
\end{abstract}

\end{frontmatter}

\section{Introduction}

A light--cone (LC) wave function is a localized (i.e.~normalizable)
stationary solution of the LC Schr\"odinger equation $i \partial_\tau |
\Psi (\tau) \ket = H_\mathrm{LC} | \Psi (\tau) \ket$, which describes
the evolution of a state $|\Psi (\tau) \ket$ in LC time $\tau \equiv
x^\p = x^0 + x^3$, conjugate to the LC Hamiltonian $H_\mathrm{LC}
\equiv P^\m = P^0 - P^3$ \cite{brodsky:98,heinzl:00}. In quantum 
field theory, a stationary solution $|\Psi \ket$ has a LC Fock
expansion which in principle does not terminate, as can be seen
e.g.~for the pion, $| \pi \ket = \psi_2 | q \bar q \ket + \psi_3 | q
\bar q g \ket + \psi_4 | q \bar q q \bar q \ket + \ldots$. The
LC wave functions are the amplitudes $\psi_n$ to find $n$ particles
with momenta $p_i$ (quarks $q$, antiquarks $\bar q$ or gluons $g$) in
a pion of momentum $P$, i.e.~$\psi_n (p_1 , \ldots p_n ; P) \equiv
\bra n |\pi \ket = \bra p_1, \ldots , p_n | \pi (P) \ket$.  We shall
see in a moment that the momentum dependence of the $\psi_n$ is of a
very peculiar nature.

Upon choosing LC rather than ordinary time one benefits from the
following virtues. First of all, the field theory vacuum becomes
`trivial', in particular $P^\m | 0 \ket = 0$. As a result,
disconnected vacuum contributions do not mix with the LC wave
functions: $\bra 0 | P^\m | n \ket = 0$. Unlike in ordinary
`instant--form' quantization, boosts become \textit{kinematical}
(interaction independent). This is closely related to a twodimensional
Galilei invariance which leads to a \textit{separation} of
center--of--mass and relative coordinates. These features imply that
the $\psi_n$ only depend on the \textit{frame independent} relative
momenta $x_i = p_i^\p/P^\p$ and $\vc{k}_{\perp i} = x_i
\vc{P}_{\!\perp} - \vc{p}_{\perp i}$.

As the number of difficulties is always conserved, the list of
advantages goes along with a list of problems. There is the conceptual
question how nontrivial condensates can arise in a trivial
vacuum. The answer is crucial for understanding spontaneous symmetry
breaking in the LC framework.  Instead of boosts, rotations and parity
become dynamical and thus complicated. There are also more
technical problems. Due to the lack of explicit covariance and
rotational invariance, renormalization becomes very difficult beyond
one loop. There are simply not enough \textit{manifest} symmetry
principles providing a guideline. As a result, there is an abuncance
of counterterms which can even become non--local ($\sim 1/k^\p$).  In
addition, LC field theories are constrained systems. This makes their
quantization nonstandard and somewhat involved. Finally, the LC
Schr\"odinger equation in general represents an infinity of coupled,
nonlinear integral equations for the amplitudes $\psi_n$. To solve
them one has to resort to truncations. It should be noted that for an
\textit{ab--initio} calculation of LC wave functions basically all
this problems have to be solved.

If it is so hard, then why should one calculate LC wave functions? It
is worth the effort because their determination implies the knowledge
of the entire hadron structure (just like the determination of the
Coulomb wave functions implies the knowledge of the hydrogen
structure). Unlike for nonrelativistic systems, however, there is a
subtlety. The (relativistic) LC wave functions depend on a
\textit{resolution scale} $Q$, $\psi_n = \psi_n (Q)$. It is well known
that there are two basic regimes, a `soft' and a `hard' one. In the
hard regime, one has $Q \gg \Lambda_\mathrm{QCD}$ so that perturbative
QCD is applicable. The LC wave functions describe the distribution of
\textit{partons}. If the resolution scale $Q$ is of the order of
$\Lambda_\mathrm{QCD}$, in the soft regime, perturbation theory no
longer works. Hadrons are believed to consist of effective
\textit{constituent} quarks which should again be reflected in the
nature of the LC wave functions. Low--energy effective field theories
seem to provide a reasonable description of the soft regime, at least
as far as spontaneous chiral symmetry breaking ($\chi$SB) is
concerned. Chiral quark models, in particular, with \textit{explicit}
quark degrees of freedom, should yield soft LC wave functions
describing hadrons as bound states of a few constituent quarks. Such a
description should be valid for $Q$ $\simleq$ 1 GeV. Whether these
expectations are true will be examined in this contribution.

\section{Formalism}

It turns out that most of the difficulties mentioned in the
introduction can be circumvented using a particular chiral quark model
originally due to Nambu and Jona-Lasinio \cite{nambu:61a}. For two
flavors the Lagrangian is
\be
  \mathcal{L} = \bar \psi (i \partial \!\!\!/ - m_0) \psi - G [(\bar
  \psi \psi)^2 + (\bar \psi i \gamma_5 \vcg{\tau} \psi)^2] \; .
\ee
The model is not renormalizable whence it requires a cutoff $\Lambda$,
the value of which is fixed by phenomenology. Beyond a critical
coupling $G_c$ one has spontaneous $\chi$SB together with dynamical
mass generation, $m_0 \to m \sim \cond$. The current quarks thus
become constituent quarks with a mass proportional to the chiral
condensate. As a constituent picture is realized one expects a
truncation of the LC Schr\"odinger equation to work reasonably
well. We avoid to solve complicated constraint equations
\cite{dietmaier:89,bentz:99,itakura:99} by using a Schwinger--Dyson
approach as pioneered by `t~Hooft \cite{thooft:74}. It consists of two
steps. First one solves the Schwinger--Dyson equation for the quark
self--energy $\Sigma$ in mean--field approximation (equivalent to the
large--$N_C$ limit). This yields the gap equation $\Sigma = const = m
\sim \cond_\Lambda$.  To make sense of the quadratically divergent
expression for the condensate one uses an \textit{invariant--mass
cutoff} $\Lambda$ satisfying $M_0^2 \equiv (k_\perp^2 + m^2)/x (1-x)
\le \Lambda^2$, $x \equiv k^\p / \Lambda$ (for an alternative, see
\cite{lenz:00}). It regulates the divergence of the condensate both
for $x \to 0 , 1$ and $k_\perp \to \infty$, and is related to the
3--vector cutoff $\Lambda_3 \ge |\vc{k}| $ by $\Lambda^2 = 4
(\Lambda_3^2 + m^2)$. Accordingly, the result for the condensate
agrees with the usual one \cite{heinzl:00,nambu:61a}.

Knowing the effective quark mass $m$ one performs the second step in
the program and solves the Bethe--Salpeter equation in the
pseudoscalar channel (using ladder approximation/large--$N_C$). The LC
wave function of the pion is obtained by three--dimensional reduction
(integrating over LC energy). In the chiral limit, the result is
\be
  \label{PI_WF}
  \left( \begin{array}{cc} 
              \psi_{2 \uparrow \uparrow} & \psi_{2 \uparrow\downarrow} \\
              \psi_{2 \downarrow\uparrow}  & \psi_{2 \downarrow\downarrow} 
              \end{array}
       \right) 
  = - \frac{N}{k_\perp^2 +
  m^2} \left( \begin{array}{cc} 
              -2m k_- & m^2 - k_\perp^2 \\
              k_\perp^2 - m^2  & - 2m k_+
              \end{array}
       \right) \theta(\Lambda^2 - M_0^2) \, , 
\ee
where we have defined $k_\pm \equiv k_1 \pm i k_2$ corresponding to
the $L_z = \pm 1$ components of the wave function.  $N$ is a
normalization constant to be determined later. The step function
implements the invariant--mass--cutoff. It is only for this cutoff
that the wave function becomes normalizable.

\section{Results}

With the pion wave function at hand one can go on and calculate
observables. To make life simple I will always work in the chiral
limit $m_0 = 0 = M_\pi$. Due to this limit the wave function
(\ref{PI_WF}) does not depend on $x$ (apart from the cutoff). 
Further simplifications arise in the large--cutoff limit
(LCL). This amounts to keeping only the leading order in $\epsilon^2
\equiv m^2 / \Lambda^2$. As a result one can find analytic expressions for all
observables\footnote{The same logic is used in the context of the
instanton model \cite{diakonov:95} where $\epsilon^2$ corresponds to
the `packing fraction' of instantons in the vacuum.}. The LC wave
function (\ref{PI_WF}) simplifies to $\psi_{2 \uparrow
\uparrow} = \psi_{2 \downarrow \downarrow} = 0$ and $\psi_{2 \uparrow
\downarrow} = - \psi_{2 \downarrow \uparrow} = N \theta (\Lambda^2 -
M_0^2)$. The nontrivial components of the wave function thus just
become step functions \cite{radyushkin:95}.  We are thus left with two
parameters, $N$ and $\Lambda$, which are determined by normalizing
$\psi_2$ to unity (\textit{enforcing} a constituent picture) and using
the pion decay constant $f_\pi$. The latter is basically given by the
wave function `at the origin' \cite{lepage:81}. This results in $N =
\sqrt{3} / f_\pi$ and $\Lambda = 4 \pi f_\pi \simeq 1.16$ GeV. The
value for $\Lambda$ exactly coincides with the Georgi--Manohar scale
\cite{manohar:84} below which chiral effective theories make sense. It
is also consistent with the constraint on the wave function stemming
from $\pi^0 \to 2 \gamma$ \cite{lepage:81},
\be
\label{ANOMALY}
  \int dx \, \psi_{2 \uparrow \downarrow} (x, \vc{0}_\perp ) =
  \sqrt{3}/f_\pi \equiv N \; .
\ee
With all parameters determined one can use the Drell--Yan formula as
discussed e.g.~in \cite{lepage:81} and calculate the pion
\textit{charge radius}. The result, $r_\pi^2 = 3 / 4 \pi^2 f_\pi^2
\simeq $ (0.60 fm)$^2$, agrees with those obtained in a covariant
Bethe--Salpeter approach \cite{blin:88} and the instanton model
\cite{diakonov:95}. The slight difference  compared to the
experimental value of 0.66 fm stems from the use of the LCL.

\begin{figure}[H]
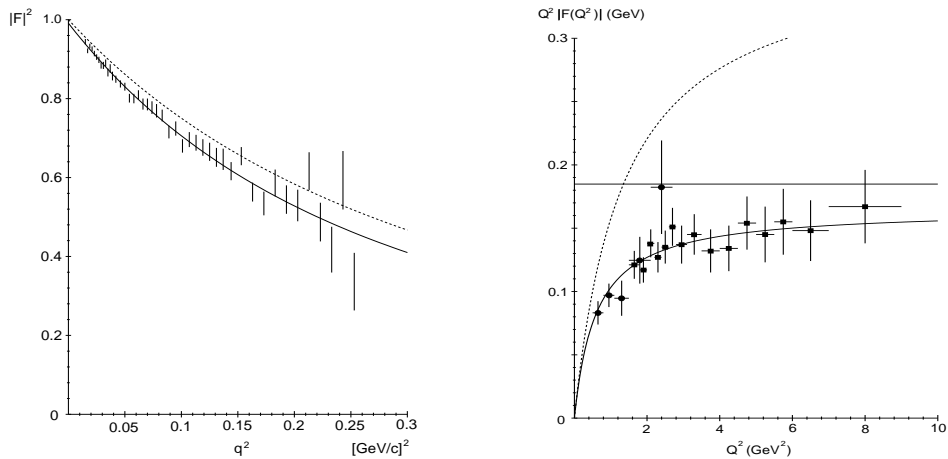

  \begin{minipage}[b]{\halfwidth}
    \centering
    \includegraphics[height=6cm,width=\medwidth]{NJL_FF.epsf}
  \end{minipage}
  \begin{minipage}[b]{\halfwidth}
    \centering
    \includegraphics[height=6cm,width=\medwidth]{pi_gamma_bw.epsf}
    \end{minipage}
\caption{{\it left:} pion electromagnetic form factor squared vs.~momentum
         transfer $q^2 \equiv q_\perp^2$; dashed line: pole fit
         $|F|^2 = 1 /(1 + q^2 r_\pi^2 / 6)$ with $r_\pi = 0.60$ fm;
         full line: same formula with experimental
         value as extracted from the data
         \protect\cite{amendolia:86};
         {\it right:} $\pi \gamma$ transition  
         form factor times momentum transfer $Q^2$; dashed line: NJL
         pole formula (\protect\ref{POLE}); full line: same formula with
         $\Lambda^2$ replaced by $\Lambda^2 / 2$; full horizontal line:
         asymptotic value $2 f_\pi$; circles: CELLO data
         \protect\cite{behrend:91}; squares: CLEO data
         \protect\cite{gronberg:98}.}
\end{figure}

A pole fit to the form factor data of Amendolia et
al. \cite{amendolia:86} is shown in the left--hand part of Fig.~1.
The momentum space wave function (\ref{PI_WF}) readily yields the
r.m.s.~transverse momentum, $\bra k_\perp^2 \ket = \Lambda^2 / 10
\simeq $ (370 MeV$^2$), which makes it obvious that the pion is a
highly relativistic system. 

The valence quark distribution $f^v$ (or pion \textit{structure
function}) is given by the square of the wave function integrated over
$k_\perp$ resulting in $f^v (x) = 6 x (1 - x)$. This coincides with
the results of \cite{bentz:99,shigetani:93} and qualitatively agrees
with the empirical parton distributions of \cite{glueck:99} if the
relevant scale is defined as $ Q^2 \equiv \bra x \ket (1 -\bra x \ket
)
\Lambda^2 \simeq$ (600 MeV)$^2$, the mean $\bra x \ket$ being 1/2.

The \textit{transition form factor} for the process $\gamma \gamma^*
\to \pi^0$ can be calculated according to \cite{lepage:81}. One finds
the pole formula
\be
\label{POLE}
  F_{\gamma \gamma^* \pi} (Q^2) = \frac{1}{4 \pi^2 f_\pi} \frac{1}{1 +
  Q^2 / \Lambda^2} \; , 
\ee
which is displayed in the right--hand part of Fig.~1. The low--$Q^2$
behavior is fixed by (\ref{ANOMALY}) and thus fine. The behavior for
$Q^2 \to \infty$ is off by a factor of two, which, however, is not
bothersome as this regime is way beyond where the NJL model makes
sense. One can fix the large--$Q^2$ behavior by introducing an
effective, spin--averaged wave function \cite{radyushkin:95}. This
simply amounts to replacing $N \to 2N$ and $\Lambda^2 \to \Lambda^2
/2$ in (\ref{PI_WF}) and (\ref{POLE}). In this case, (\ref{ANOMALY})
cannot be maintained and the logic based on low--energy chiral
dynamics does no longer apply.

The \textit{pion distribution amplitude} $\phi_\mathrm{NJL}$ is given
by the $k_\perp$--integral of the wave function. As the square of a
step function is again a step function the distribution amplitude
coincides with $f^v$ and hence with the \textit{asymptotic}
distribution amplitude, $\phi_\mathrm{NJL} (x) = 6 x (1-x) \equiv
\phi_\mathrm{as} (x)$.  Recently this quantity has been measured
at Fermilab \cite{ashery:99}. At a rather low momentum scale of $Q
\simeq$ 3 GeV $\phi$ turns out to be close to asymptotic.  This
provides further evidence that the approach presented above makes
sense.

\section{Conclusions}

I have analytically determined the LC wave function of the pion within
a simple field theoretic model (NJL) which is known to yield a good
description of $\chi$SB. With the pion wave function at hand I was
able to calculate a number of observables which were accurate to
within 10 \%, consistent with the estimated limitations of the LCL
employed. Particularly important is the result that a constituent
picture does make sense: higher Fock states seem to be unimportant. If
one wants to go beyond the LCL, one faces a fine--tuning problem in
$\epsilon^2$.  Delicate numerical fits will thus be necessary.  It
would be interesting to refine the presented method by choosing more
realistic models. Possibilities would be (i) the instanton vacuum
\cite{diakonov:95} which is more closely related to QCD, (ii) an
elaborate Schwinger--Dyson approach with more realistic kernels and
propagators \cite{maris:00}, or (iii) an effective field theory
approach in terms of constituent quarks which, at the moment, is still
in its infancy \cite{manohar:84}.

\section*{Acknowledgments}
I cordially thank the organizers S.~Bielefeld, L.~Hollenberg and
H.-C.~Pauli for their efforts which resulted in such a stimulating
meeting.

%\bibliographystyle{elsart-num}
%\bibliography{../../bibfiles/heinzl}

%\end{thebibliography}

\end{document}